\documentclass[12pt,reqno,a4paper]{amsart}
\usepackage{amsmath,amssymb,dsfont,graphicx,cite}
\def\beq{\begin{equation}}
\def\eeq{\end{equation}}
\def\bea{\begin{eqnarray}}
\def\eea{\end{eqnarray}}

\newtheorem{prop}{Proposition}
\newtheorem{lemma}{Lemma}

\newtheorem{corollary}{Corollary}
\makeatletter
\expandafter\let\expandafter
\reset@font\csname reset@font\endcsname
\def\subeqnarray{\arraycolsep1pt
    \def\@eqnnum\stepcounter##1{\stepcounter{subequation}
        {\reset@font\rm(\theequation\alph{subequation})}}
\jot5mm     \eqnarray}

\makeatother
\newcommand{\bbZ}{{\mathbb Z}}
\newcommand{\bbR}{{\mathbb R}}
\def\ep{\epsilon}

\def\endpf{\begin{flushright}$\square$\end{flushright}}

\def\su2{{\mathfrak {su}}(2)}
\def\e3{{\mathfrak {e}}(3)}

\begin{document}
%%%%%%%%%%%%%%%%%%%%%%%%%%%%%%%
%%%%%%%%%%%%%%%%%%%%%%%%%%%%%%%
\title[On the Hamiltonian structure of HK discretization
of the Euler top]
{\bf On the Hamiltonian structure  of
Hirota-Kimura discretization of the Euler top}
%%%%%%%%%%%%%%%%%%%%%%%%%%%%%%%
%%%%%%%%%%%%%%%%%%%%%%%%%%%%%%%
\author[MATTEO PETRERA and YURI B. SURIS]
{MATTEO PETRERA${}^{\dag}$ and YURI B. SURIS${}^{\diamond}$}

\thanks{${}^\dag$ \texttt{petrera@ma.tum.de}}
\thanks{${}^\diamond$ \texttt{suris@ma.tum.de}}

\maketitle

\centerline{\it Zentrum Mathematik,
Technische Universit\"at M\"unchen}
\centerline{\it Boltzmannstr. 3, D-85747 Garching bei M\"unchen,
Germany}

%%%%%%%%%%%%%%%%%%%%%%%%%%%%%%%
%%%%%%%%%%%%%%%%%%%%%%%%%%%%%%%
\begin{abstract}

This paper deals with a remarkable integrable discretization of
the $so(3)$ Euler top introduced by Hirota and Kimura. Such a
discretization leads to an  explicit map, whose integrability has
been understood by finding two independent integrals of motion and
a solution in terms of elliptic functions. Our goal is the
construction of its Hamiltonian formulation. After giving a
simplified and streamlined presentation of their results, we
provide a bi-Hamiltonian structure for this discretization, thus
proving its integrability in the standard Liouville-Arnold sense.

\end{abstract}
%%%%%%%%%%%%%%%%%%%%%%%%%%%%%%%
%%%%%%%%%%%%%%%%%%%%%%%%%%%%%%%
%\maketitle
%\tableofcontents
%%%%%%%%%%%%%%%%%%%%%%%%%%%%%%%
%%%%%%%%%%%%%%%%%%%%%%%%%%%%%%%
\section{Introduction} \label{sect: intro}
%%%%%%%%%%%%%%%%%%%%%%%%%%%%%%%
%%%%%%%%%%%%%%%%%%%%%%%%%%%%%%%

This paper deals with a remarkable integrable discretization for
one of the basic integrable systems, the three-dimensional Euler
top, which describes the motion of the free rigid body with a
fixed point. Equations of motion of the Euler top in the body
frame read
\begin{equation}\label{E x}
\dot{x}_1=\alpha_1 x_2 x_3,\quad \dot{x}_2=\alpha_2 x_3 x_1, \quad
\dot{x}_3=\alpha_3 x_1x_2.
\end{equation}
where $x = (x_1,x_2,x_3)\in \mathbb{R}^3$, and the real
coefficients $\alpha_i$ are parameters of the system. We will
denote the vector of parameters by $\alpha =
(\alpha_1,\alpha_2,\alpha_3) \in \mathbb{R}^3$. Throughout this
paper we will use an abbreviated notation, according to which
$(ijk)$ stands for any cyclic permutation of $(123)$. Thus, system
(\ref{E x}) takes with this notation the form
\begin{equation}\label{E}
\dot{x}_i=\alpha_i x_j x_k.
\end{equation}
The coordinates $x_i$ stand either for the angular velocities
$\Omega_i$, in which case the coefficients $\alpha_i$ are given by
\begin{equation}\label{alpha for omega}
\alpha_i = \frac{I_j-I_k}{I_i},
\end{equation}
or otherwise for the angular momenta $M_i$, in which case the
coefficients $\alpha_i$ are given by
\begin{equation}\label{alpha for M}
\alpha_i = \frac{1}{I_k}-\frac{1}{I_j}.
\end{equation}
Here $I_i$ are the principal moments of inertia of the body. The
relation between the two formulations is given by
$M_i=I_i\Omega_i$. Integrability features of the Euler top include
\cite{S,A,RSTS}: a bi-Hamiltonian structure, i.e. the existence of
two compatible invariant Poisson structures on the phase space;
two independent integrals of motion, which are in involution with
respect to any of the invariant Poisson brackets; a Lax
representation; explicit solutions in terms of elliptic functions.
For the reader's convenience, some of these features are briefly
exposed in Sect. \ref{sect: cont Euler top}.

The general problem of integrable discretization of integrable
systems is dealt with in the monograph \cite{S}. One finds there
also a detailed exposition of an integrable discretization of the
Euler top, due to Veselov and Moser \cite{V, MV}. The basic
feature of this discretization is that it comes from a discrete
Lagrangian formulation on the Lie group $SO(3)$. Upon a reduction
to $so(3)^*$, it produces a correspondence, i.e. a multi-valued
map, each branch of which is Poisson with respect to the
Lie-Poisson bracket on $so(3)^*$, like the original phase flow.
Moreover, it shares the integrals of motion and the Lax
representation with the original continuous time flow. This Lax
representation is related to matrix factorizations.

A class of discretizations of the Euler top sharing the integrals
of motion with the continuous system has been introduced and
studied in \cite{BLS}. These discretizations are characterized by
the equations of motion
\begin{equation}\label{dE BLS}
\tilde{x}_i-x_i=\gamma\alpha_i(\tilde{x}_j+x_j)(\tilde{x}_k+x_k).
\end{equation}
Here and below tilde denotes the shift $t\mapsto t+\epsilon$ in
the discrete time $\epsilon\mathbb Z$, where $\epsilon$ is a
(small) time step. In other words, in Eq. (\ref{dE BLS}) (and in
similar situation throughout the paper) we consider $x_i$ as
functions on $\epsilon\bbZ$, and we write $x_i$ for $x_i(n\ep)$
and $\tilde x_i$ for $x_i(n\epsilon +\epsilon)$, $n\in
\mathbb{Z}$. In Eq. (\ref{dE BLS}), it is assumed that
$\gamma\sim\epsilon/4$ is some real-valued function on the phase
space. Then the map (\ref{dE BLS}) approximates, for small
$\epsilon$, the time $\epsilon$ shift along the trajectories of
the continuous flow (\ref{E x}). In \cite{BLS} the functions
$\gamma$ have been characterized for which the map
$(x_1,x_2,x_3)\mapsto(\tilde{x}_1,\tilde{x}_2,\tilde{x}_3)$
defined by Eq. (\ref{dE BLS}) shares the invariant Poisson
structure with the continuous system. In particular, the function
$\gamma$ for the Veselov-Moser discretization has been determined.
A further integrable discretization of the Euler top belonging to
the family (\ref{dE BLS}) was proposed in \cite{F}. Interestingly,
the simplest choice $\gamma=\epsilon/4$ leads to a a map which
{\em does not} preserve the original Poisson structure.
Discretizations (\ref{dE BLS}) share the Lax matrix with the
continuous time Euler top. They are {\it implicit}, since these
formulas represent a system of algebraic (nonlinear) equations for
$(\tilde{x}_1,\tilde{x}_2,\tilde{x}_3)$ which does not possess a
simple closed-form solution.

The present paper deals with the following beautiful {\em
explicit} discretization of equations of motion (\ref{E}),
introduced by Hirota and Kimura \cite{HK}:
\begin{equation}\label{dE x}
\tilde{x}_i-x_i=\delta_i(\tilde{x}_jx_k+x_j\tilde{x}_k).
\end{equation}
Here one can take
\begin{equation}
\label{delta} \delta_i =  \frac{\epsilon\alpha_i}{2},
\end{equation}
we will adopt this choice for the vector of parameters $\delta =
(\delta_1,\delta_2,\delta_3) \in \mathbb{R}^3$ throughout the
paper. This discretization is explicit, since the algebraic
equations (\ref{dE x}) are linear with respect to
$(\tilde{x}_1,\tilde{x}_2,\tilde{x}_3)$, and thus they can be
solved in a closed form (see Sect. \ref{sect: Hirota Euler} for
further details). Hirota and Kimura presented some of the
integrability attributes for their discretization: two independent
integrals of motion and a solution in terms of elliptic functions.
Other attributes, like the Hamiltonian formulation and the Lax
representation, has not been mentioned by them. The main goal of
the present paper is to fill the first of these two gaps by
providing a bi-Hamiltonian structure for the Hirota-Kimura
discretization, and thereby to prove its integrability in the
standard Liouville-Arnold sense.

We found it worthwhile to give also a simplified and streamlined
presentation of the results found in \cite{HK}. Indeed, the
discretization of the Lagrange top given by Kimura and Hirota
later in \cite{KH}, as well as some preliminary results by Ratiu
\cite{R}, indicate that the map (\ref{dE x}) might be just a tip
of an iceberg, a huge collection of discretizations of integrable
systems of classical mechanics. We plan to develop this topic in a
series of upcoming publications.

It is an established fact that many of the most important
integrable systems can be found in the classical literature on
differential geometry. Usually this refers to solitonic partial
differential equations, like the sine-Gordon equation, but it
turns out to be true also for the integrable map (\ref{dE x}): a
1951 paper in ``Mathematische Nachrichten'' by H.\ Jonas is
devoted to a birational map
$(x,y,z)\mapsto(\tilde{x},\tilde{y},\tilde{z})$ given by
\begin{equation}\label{Jonas}
x+\tilde{x}+y\tilde{z}+z\tilde{y}=0,\quad
y+\tilde{y}+z\tilde{x}+x\tilde{z}=0,\quad
z+\tilde{z}+x\tilde{y}+y\tilde{x}=0,
\end{equation}
which differs only unessentially from (\ref{dE x}). The map
(\ref{Jonas}) has an origin in the spherical geometry, $(x,y,z)$
and $(\tilde{x},\tilde{y},\tilde{z})$ being the cosines of the
side lengths of two spherical triangles with complementary angles.
Jonas' results include integrals of the map (\ref{Jonas}) and its
solution in terms of elliptic functions. Thus, \cite{J} seems to
be one of the earliest precursors of the theory of integrable
maps.

%%%%%%%%%%%%%%%%%%%%%%%%%%%%%%%
%%%%%%%%%%%%%%%%%%%%%%%%%%%%%%%
\section{Euler top}
\label{sect: cont Euler top}
%%%%%%%%%%%%%%%%%%%%%%%%%%%%%%%
%%%%%%%%%%%%%%%%%%%%%%%%%%%%%%%

The aim of this Section is to recall some of the main features of
the integrable continuous-time Hamiltonian flow (\ref{E}).
\begin{prop}\label{th: ET H}
Let $\beta = (\beta_1,\beta_2, \beta_3) \in \mathbb{R}^3$ be a
constant vector. A quadratic function
\begin{equation}\label{E H}
H^{(\beta)} = \frac{1}{2}(\beta_1x_1^2+\beta_2x_2^2+\beta_3x_3^2)
\end{equation}
is an integral of motion for (\ref{E}) if and only if $\beta \perp
\alpha$, i.e. if
$\beta_1\alpha_1+\beta_2\alpha_2+\beta_3\alpha_3=0$.
\end{prop}
{\bf Proof:} An easy computation based on Eq. (\ref{E}) shows that
\[
\frac{d}{dt}H^{(\beta)}=
(\beta_1\alpha_1+\beta_2\alpha_2+\beta_3\alpha_3)x_1x_2x_3.
\]
\endpf

Since the orthogonal complement of the vector $\alpha$ is
two-dimensional, there are two independent integrals of motion. It
is sometimes convenient to use a special basis of the orthogonal
complement just mentioned, consisting of vectors with one
vanishing component.

\begin{corollary}
The three quadratic functions
\begin{equation}\label{G}
G_i = \frac{1}{2}(\alpha_jx_k^2-\alpha_kx_j^2)
\end{equation}
are integrals of motion for (\ref{E}). Of course, only two of them
are (linearly) independent since
$\alpha_1G_1+\alpha_2G_2+\alpha_3G_3=0$.
\end{corollary}

Notice that any function $H^{(\beta)}$ is a linear combination of
the $G_i$'s:
$$
\alpha_i H^{(\beta)} = \beta_j G_k -\beta_k G_j.
$$

In the angular velocities formulation, a basis of the orthogonal
complement $\alpha^\perp$ can be chosen consisting of $\beta^{(1)}
=(I_1,I_2,I_3)$ and $\beta^{(2)}=(I_1^2,I_2^2,I_3^2)$. In the
angular momenta formulation, a basis of $\alpha^\perp$ consists of
$\beta^{(1)} =(1/I_1,1/I_2,1/I_3)$ and $\beta^{(2)}= (1,1,1)$.

\begin{prop}\label{th: E ham}
Let $\beta \perp \alpha$, and let $\gamma =
(\gamma_1,\gamma_2,\gamma_3)\in \mathbb R^3$ satisfy
\begin{equation}\label{alpha=beta x gamma}
\alpha_i=\beta_j\gamma_k-\beta_k\gamma_j,
\end{equation}
so that $\gamma \perp \alpha$. Then the system (\ref{E}) is
Hamiltonian with the Hamilton function $H^{(\beta)}$ with respect
to the Poisson bracket
\begin{equation}
\left\{x_i,x_j\right\}^{(\gamma)}= \gamma_kx_k.
\end{equation}
\end{prop}
{\bf Proof:} A direct verification:
\begin{eqnarray*}
\{x_i,H^{(\beta)}\}^{(\gamma)}
 & = & \beta_jx_j \{x_i,x_j\}^{(\gamma)}+
 \beta_kx_k\{x_i,x_k\}^{(\gamma)}=\\
 & = & (\beta_j\gamma_k-\beta_k\gamma_j) x_jx_k =\alpha_ix_jx_k.
\end{eqnarray*}
\endpf

Propositions \ref{th: ET H} and \ref{th: E ham} show the
bi-Hamiltonian property of the Euler top. Referring to the angular
velocities the system has two Hamiltonian formulations:
\[
H = \frac{1}{2}(I_1\Omega_1^2+I_2\Omega_2^2+I_3\Omega_3^2)\qquad
{\rm with} \qquad \left\{\Omega_i,\Omega_j\right\}=
\frac{I_k}{I_iI_j}\Omega_k,
\]
and
\[
H=\frac{1}{2}(I_1^2\Omega_1^2+I_2^2\Omega_2^2+I_3^2\Omega_3^2)\qquad
{\rm with} \qquad \left\{\Omega_i,\Omega_j\right\}=
\frac{1}{I_iI_j}\Omega_k.
\]
Referring to the angular momenta, the system also has two
Hamiltonian formulations:
\[
H=\frac{1}{2}\left(\frac{M_1^2}{I_1}+\frac{M_2^2}{I_2}+
\frac{M_3^2}{I_3}\right)\qquad {\rm with} \qquad
\left\{M_i,M_j\right\}=M_k,
\]
and
\[
H=\frac{1}{2}(M_1^2+M_2^2+M_3^2)\qquad  {\rm with} \qquad
\left\{M_i,M_j\right\}= \frac{1}{I_k}M_k.
\]

%%%%%%%%%%%%%%%%%%%%%%%%%%%%%%%
%%%%%%%%%%%%%%%%%%%%%%%%%%%%%%%
\section{Hirota-Kimura discretization of the Euler top}
\label{sect: Hirota Euler}
%%%%%%%%%%%%%%%%%%%%%%%%%%%%%%%
%%%%%%%%%%%%%%%%%%%%%%%%%%%%%%%

We now turn to the study of the map (\ref{dE x}). Though the
vector of parameters $\delta$ is arbitrary, we will think of it as
related to $\alpha$ as in Eq. (\ref{delta}).

%%%%%%%%%%%%%%%%%%%%%%%%%%%%%%%
%%%%%%%%%%%%%%%%%%%%%%%%%%%%%%%
\subsection{Integrals of motion}
\label{subsect: Hirota Euler integrals}
%%%%%%%%%%%%%%%%%%%%%%%%%%%%%%%
%%%%%%%%%%%%%%%%%%%%%%%%%%%%%%%

An explicit form of this map can be easily obtained. Considering
Eq. (\ref{dE x}) as a system of linear equations for the updated
variables $\tilde{x}_i$, one finds immediately its solution:
\begin{equation}
\nonumber
\begin{pmatrix} \tilde{x}_1 \\ \tilde{x}_2 \\ \tilde{x}_3
\end{pmatrix}=
\begin{pmatrix} 1 & -\delta_1 x_3 & -\delta_1 x_2 \\
-\delta_2 x_3 & 1 & -\delta_2 x_1 \\
-\delta_3 x_2 & -\delta_3 x_1 & 1 \end{pmatrix}^{-1}
\begin{pmatrix} x_1 \\ x_2 \\ x_3 \end{pmatrix}.
\end{equation}
Note also that, considering Eq. (\ref{dE x}) as a system of linear
equations for $\tilde{x}_i$, one finds the alternative formula
\begin{equation} \nonumber
\begin{pmatrix} \tilde{x}_1 \\ \tilde{x}_2 \\ \tilde{x}_3
\end{pmatrix}=
\begin{pmatrix} 1 & \delta_1 \tilde{x}_3 & \delta_1 \tilde{x}_2 \\
\delta_2 \tilde{x}_3 & 1 & \delta_2 \tilde{x}_1 \\
\delta_3 \tilde{x}_2 & \delta_3 \tilde{x}_1 & 1
\end{pmatrix} \begin{pmatrix} x_1 \\ x_2 \\ x_3 \end{pmatrix}.
\end{equation}

We will use the notation
\begin{equation} \nonumber
A(x,\delta) =  \begin{pmatrix} 1 & -\delta_1 x_3 & -\delta_1 x_2 \\
-\delta_2 x_3 & 1 & -\delta_2 x_1 \\
-\delta_3 x_2 & -\delta_3 x_1 & 1
\end{pmatrix} ,
\end{equation}
so that the equations of the map can be written as
\[
\tilde{x}= A^{-1}(x,\delta)x=A(\tilde{x},-\delta)x.
\]

\begin{prop}\label{th: dE F}
The quantities
\begin{equation}\label{dE F}
F_i = \frac{1-\delta_k\delta_ix_j^2}{1-\delta_i\delta_jx_k^2},
\end{equation}
are integrals of motion for the map (\ref{dE x}). Of course, there
are only two independent integrals since $F_1F_2F_3=1$.
\end{prop}

{\bf Proof:} Equation $\tilde{F}_i=F_i$ can be re-written as
\[
(1-\delta_k\delta_i\tilde{x}_j^2)(1-\delta_i\delta_jx_k^2)=
(1-\delta_i\delta_j\tilde{x}_k^2)(1-\delta_k\delta_ix_j^2),
\]
which is equivalent to
\[
\delta_j(\tilde{x}_k^2-x_k^2)- \delta_k(\tilde{x}_j^2-x_j^2)
=\delta_i\delta_j\delta_k(\tilde{x}_j^2x_k^2-x_j^2\tilde{x}_k^2),
\]
that is, to
\[
\delta_j(\tilde{x}_k+x_k)(\tilde{x}_k-x_k)-
\delta_k(\tilde{x}_j+x_j)(\tilde{x}_j-x_j)=
\delta_i\delta_j\delta_k(\tilde{x}_jx_k+x_j\tilde{x}_k)
(\tilde{x}_kx_j-x_k\tilde{x}_j).
\]
Using the equations of motion (\ref{dE x}) on both sides of the
latter formula, we arrive at
\[
(\tilde{x}_k+x_k)(\tilde{x}_ix_j+x_i\tilde{x}_j)-
(\tilde{x}_j+x_j)(\tilde{x}_kx_i+x_k\tilde{x}_i)
=(\tilde{x}_i-x_i)(\tilde{x}_kx_j-x_k\tilde{x}_j),
\]
which is an algebraic identity. \endpf

The relation between $F_i$'s and the integrals of the continuous
time Euler top is straightforward:
\begin{equation}
F_i=1+\frac{\epsilon^2\alpha_i}{4}\,G_i+O(\epsilon^4).  \nonumber
\end{equation}

\begin{corollary}\label{th: dE H}
Let $\beta\perp\delta$. Then the following three functions are
integrals of motion for the map (\ref{dE x}):
\begin{equation}
H_i^{(\beta)}=\frac{H^{(\beta)}}{1-\delta_j\delta_kx_i^2},  \nonumber
\end{equation}
where the common numerator $H^{(\beta)}$ is an integral of the
continuous time Euler top given in Eq. (\ref{E H}).
\end{corollary}

{\bf Proof:} We show that $H_i^{(\beta)}$ can be expressed in terms
of the $F_i$'s given in Eq. (\ref{dE F}):
\begin{eqnarray*}
\delta_i H_i^{(\beta)} & = &
\frac{-(\beta_j\delta_j+\beta_k\delta_k)x_i^2+
\beta_j\delta_ix_j^2+\beta_k\delta_ix_k^2}
{1-\delta_j\delta_kx_i^2}\\
& = & \frac{\beta_j(\delta_ix_j^2-\delta_jx_i^2)+
\beta_k(\delta_ix_k^2-\delta_kx_i^2)}
{1-\delta_j\delta_kx_i^2}\\
& = & \frac{\beta_j}{\delta_k}
\left(1-\frac{1-\delta_k\delta_ix_j^2}{1-\delta_j\delta_kx_i^2}
\right) + \frac{\beta_k}{\delta_j}
\left(1-\frac{1-\delta_1\delta_jx_k^2}{1-\delta_j\delta_kx_i^2} \right)\\
& = & \frac{\beta_j}{\delta_k}\left(1-\frac{1}{F_k}\right)+
\frac{\beta_k}{\delta_j}(1-F_j).
\end{eqnarray*}

\endpf

%%%%%%%%%%%%%%%%%%%%%%%%%%%%%%%
%%%%%%%%%%%%%%%%%%%%%%%%%%%%%%%
\subsection{Invariant volume form}
\label{subsect: Hirota Euler volume}
%%%%%%%%%%%%%%%%%%%%%%%%%%%%%%%
%%%%%%%%%%%%%%%%%%%%%%%%%%%%%%%

Next, we establish the existence of an invariant measure for  the
map (\ref{dE x}). Let us first give the following useful Lemma.

\begin{lemma} \label{lemma1}
For the map (\ref{dE x}) the following holds:
\begin{equation}\label{lk}
  \frac{\tilde{x}_i-\delta_i\tilde{x}_j\tilde{x}_k}
  {1-\delta_j\delta_k\tilde{x}_i^2} =
  \frac{x_i+\delta_ix_jx_k}{1-\delta_j\delta_kx_i^2},
\end{equation}
and, as a corollary,
\begin{equation}\label{minors}
\frac{(\tilde{x}_i-\delta_i\tilde{x}_j\tilde{x}_k)^2}
{(1-\delta_i\delta_k\tilde{x}_j^2)(1-\delta_i\delta_j\tilde{x}_k^2)}
=\frac{(x_i+\delta_ix_jx_k)^2}
{(1-\delta_i\delta_kx_j^2)(1-\delta_i\delta_jx_k^2)}.
\end{equation}
\end{lemma}
{\bf Proof:} We prove, for instance, Eq. (\ref{lk}). It is
equivalent to
\[
(\tilde{x}_i-\delta_i\tilde{x}_j\tilde{x}_k)(1-\delta_j\delta_kx_i^2)=
(x_i+\delta_ix_jx_k)(1-\delta_j\delta_k\tilde{x}_i^2) ,
\]
or to
\[
\tilde{x}_i-x_i-\delta_i\tilde{x}_j\tilde{x}_k-\delta_ix_jx_k=
-\delta_j\delta_kx_i\tilde{x}_i(\tilde{x}_i-x_i)
-\delta_i\delta_j\delta_k(x_i^2\tilde{x}_j\tilde{x}_k
+\tilde{x}_i^2x_jx_k).
\]
Upon using equations of motion (\ref{dE x}) on both sides of the
latter formula, we find that it is equivalent to
\[
(\tilde{x}_j-x_j)(\tilde{x}_k-x_k)=
\delta_j\delta_k(x_i\tilde{x}_j+\tilde{x}_ix_j)
(x_i\tilde{x}_k+\tilde{x}_ix_k),
\]
which is a direct consequence of Eq. (\ref{dE x}).
\endpf

Now we are in the position to prove the following claim.

\begin{prop}\label{th: inv meas}
There holds:
\begin{equation}
\det\frac{\partial \tilde{x}}{\partial x}=
\frac{\phi(\tilde{x})}{\phi(x)}, \nonumber
\end{equation}
where $\phi(x)$ is any of the functions
\begin{eqnarray}
 && \phi(x) =
 (1-\delta_i\delta_j x_k ^2)(1-\delta_j \delta_k x_i ^2),
 \label{phi1}\\
 && \phi(x) =  (1-\delta_i\delta_jx_k^2)^2
 \label{phi4}.
\end{eqnarray}
(The ratio of any two different functions $\phi(x)$ is an integral
of motion for (\ref{dE x}) due to Proposition \ref{th: dE F}).
Equivalently, the three-form
\begin{equation}
\label{vol}
    \Omega= \frac{1}{\phi(x)}\,dx_1\wedge dx_2\wedge dx_3
\end{equation}
is invariant under the map (\ref{dE x}).
\end{prop}

{\bf Proof:} First of all, we derive the following formula for the
Jacobian of the map  (\ref{dE x}):
\begin{equation}\label{Jac Deltas}
\det\frac{\partial \tilde{x}}{\partial x}=
\frac{\det A(\tilde{x},-\delta)}{\det A(x,\delta)}.
\end{equation}
Indeed, differentiating Eq. (\ref{dE x}) with respect to
$x_1,x_2,x_3$, one obtains the columns of the matrix equation
\[
\begin{pmatrix} 1 & -\delta_1 x_3 & -\delta_1 x_2 \\
-\delta_2 x_3 & 1 & -\delta_2 x_1 \\
-\delta_3 x_2 & -\delta_3 x_1 & 1 \end{pmatrix}
\frac{\partial \tilde{x}}{\partial x}=
\begin{pmatrix} 1 & \delta_1 \tilde{x}_3 & \delta_1 \tilde{x}_2 \\
\delta_2 \tilde{x}_3 & 1 & \delta_2 \tilde{x}_1 \\
\delta_3 \tilde{x}_2 & \delta_3 \tilde{x}_1 & 1
\end{pmatrix}.
\]
Computing determinants leads to Eq. (\ref{Jac Deltas}), which can
be written in length as
\begin{eqnarray}
\det \frac{\partial \tilde{x}}{\partial x} & = &
\frac{1-\delta_j\delta_k\tilde{x}_i^2-\delta_i\delta_k\tilde{x}_j^2-
\delta_i\delta_j\tilde{x}_k^2
+2\delta_i \delta_j\delta_k\tilde{x}_i\tilde{x}_j\tilde{x}_k}
     {1-\delta_j\delta_k x_i^2-\delta_i\delta_kx_j^2-\delta_i\delta_jx_k^2
-2\delta_i\delta_j\delta_kx_ix_jx_k}\nonumber\\
 & = &
 \frac{(1-\delta_i\delta_k\tilde{x}_j^2)(1-\delta_i \delta_j\tilde{x}_k^2)
-\delta_j\delta_k(\tilde{x}_i-\delta_i\tilde{x}_j\tilde{x}_k)^2}
     {(1-\delta_i\delta_kx_j^2)(1-\delta_i\delta_jx_k^2)
-\delta_j\delta_k(x_i+\delta_ix_jx_k)^2}.  \nonumber
\end{eqnarray}
Now the claim of Proposition with $\phi$ as in Eq. (\ref{phi1}),
say, follows from Eq. (\ref{minors}).
\endpf

%%%%%%%%%%%%%%%%%%%%%%%%%%%%%%%
%%%%%%%%%%%%%%%%%%%%%%%%%%%%%%%
\subsection{Invariant Poisson structure}
\label{sect: Hirota Euler Poisson}
%%%%%%%%%%%%%%%%%%%%%%%%%%%%%%%
%%%%%%%%%%%%%%%%%%%%%%%%%%%%%%%

In the construction of an  invariant Poisson structure for the map
(\ref{dE x}) we shall use the following results from \cite{BHQ}
(Proposition 15 and Corollary 16 there).

Let $f:M\to M$ be a smooth mapping of an $n$-dimensional manifold
$M$, and let $\Omega$ be a volume form invariant under $f$, i.e.,
$f^*\Omega=\Omega$. Define $\omega$ to be the dual $n$-vector
field to $\Omega$ such that $\omega\lrcorner\,\Omega=1$. Here the
symbol $\lrcorner$ denotes the contraction between multi-vector
fields and forms. If $I_1,\ldots,I_{n-2}$ are integrals of $f$
with $dI_1\wedge\cdots\wedge dI_{n-2}\neq 0$, then the bi-vector
field $\sigma = \omega\lrcorner\, dI_1\cdots\lrcorner\, dI_{n-2}$
is an invariant Poisson structure for $f$. If $J_1,\ldots,J_{n-2}$
is another set of independent integrals and $\tau =
\omega\lrcorner\, dJ_1\cdots\lrcorner\, dJ_{n-2}$ is the
corresponding Poisson structure, then $\sigma$ and $\tau$ are
compatible, i.e., for any constants $a$, $b$, the bi-vector field
$a\sigma+b\tau$ is a Poisson structure, again.

In particular, for $n=3$, if a three-form (\ref{vol}) is invariant
under $f$, so that the dual tri-vector field is given by
\[
\omega = \phi(x)\frac{\partial}{\partial x_1}\wedge
 \frac{\partial}{\partial x_2}\wedge
 \frac{\partial}{\partial x_3}\,,
\]
then for any integral $I$ of $f$ the bi-vector field
\begin{equation}\label{inv pois tensor}
\sigma = \omega  \lrcorner dI =\phi(x)\left(\frac{\partial
I}{\partial x_3}\frac{\partial}{\partial
x_1}\wedge\frac{\partial}{\partial x_2}+ \frac{\partial
I}{\partial x_1}\frac{\partial}{\partial
x_2}\wedge\frac{\partial}{\partial x_3}+\frac{\partial I}{\partial
x_2}\frac{\partial}{\partial x_3}\wedge\frac{\partial}{\partial
x_1}\right)
\end{equation}
is an invariant Poisson structure for $f$, as well as any linear
combination of such bi-vector fields. The Poisson brackets of
coordinate functions are given by
\begin{equation}\label{inv pb}
\left\{x_i,x_j\right\}= \phi(x)\frac{\partial I}{\partial x_k}\,.
\end{equation}
Applying this result to the integrals $\log F_1$, $\log F_2$,
$\log F_3$, with the three volume densities  (\ref{phi1}),
we arrive at the following statement.
\begin{prop}
The following brackets give compatible invariant polynomial
Poisson structures for the map (\ref{dE x}):
\begin{equation}\label{pb}
\left\{x_i,x_j\right\} = C_i \delta_j x_k (1-\delta_k \delta_i x_j
^2)-C_j \delta_ix_k(1-\delta_j\delta_kx_i^2),
\end{equation}
where $C_1$, $C_2$, $C_3$ are arbitrary constants.
\end{prop}

Notice that the Poisson brackets (\ref{pb}) yield three compatible
polynomial Poisson structures. Indeed, setting $C_2=C_3=0$ and
$C_1=1$, we get
\begin{equation}\label{pb1}
 \{x_1,x_2\}_1 = \delta_2x_3(1-\delta_3 \delta_1 x_2^2), \quad
 \{x_2,x_3\}_1 = 0, \quad
 \{x_3,x_1\}_1=-\delta_3x_2(1-\delta_1\delta_2x_3^2),
\end{equation}
setting $C_1=C_3=0$ and $C_2=1$, we get
\begin{equation}\label{pb2}
 \{x_1,x_2\}_2 =-\delta_1 x_3(1-\delta_2\delta_3x_1 ^2), \quad
 \{x_2,x_3\}_2 = \delta_3x_1(1-\delta_1\delta_2x_3^2), \quad
 \{x_3,x_1\}_2= 0,
\end{equation}
while setting $C_1=C_2=0$ and $C_3=0$, we get
\begin{equation}\label{pb3}
 \{x_1,x_2\}_3 =0, \quad
 \{x_2,x_3\}_3 =-\delta_2x_1(1-\delta_3\delta_1x_2^2), \quad
 \{x_3,x_1\}_3 =\delta_1x_2(1-\delta_2\delta_3x_1^2).
\end{equation}
It is easy to verify that the brackets (\ref{pb1}), (\ref{pb2}),
(\ref{pb3}) admit as Casimir functions the integrals $F_1$, $F_2$,
$F_3$,  respectively.

In the continuous limit $\epsilon\to 0$ these three brackets
correspond to the invariant linear brackets
$\{\cdot,\cdot\}^{(\gamma)}$ of the Euler top, given in
Proposition \ref{th: E ham}, with $\gamma=(0,-\alpha_3,\alpha_2)$,
$\gamma=(\alpha_3,0,-\alpha_1)$, and
$\gamma=(-\alpha_2,\alpha_1,0)$, respectively. Clearly, these
three linear brackets are linearly dependent. On the contrary, the
three polynomial brackets (\ref{pb1}), (\ref{pb2}), (\ref{pb3})
are linearly independent, if one considers linear combinations
with scalar coefficients. However, they become linearly dependent,
if one considers more general linear combinations. Indeed, the
volume density $\phi$ in Eq. (\ref{inv pois tensor}) can be
multiplied by an arbitrary integral without violating the Poisson
property. Thus, in formulating the compatibility property of such
Poisson tensors it is natural to consider their linear
combinations with coefficients being integrals of motion rather
than just numbers. In particular, the linear combination of the
brackets (\ref{pb1}), (\ref{pb2}), (\ref{pb3}) with the
coefficients
$$
(C_i,C_j,C_k)=\left(\frac{\delta_i}{F_j},\delta_jF_i,\delta_k\right)
$$
vanishes, so that there are only two independent brackets among
them.

%there holds the following simple but not very well known
%statement.
%\begin{lemma}
%Let $P$ be a Poisson tensor on $\bbR^3$, and let
%$\phi:\bbR^3\to\bbR$ be any real-valued function. Then $Q=\phi P$
%is again a Poisson tensor on $\bbR^3$.
%\end{lemma}
%{\bf Proof.} To verify the Jacobi identity for the bracket
%corresponding to $Q$, we compute:
%\begin{eqnarray*}
%\lefteqn{
%\{\{x_1,x_2\}_Q,x_3\}_Q+\{\{x_3,x_1\}_Q,x_2\}_Q+\{\{x_2,x_3\}_Q,x_1\}_Q}\\
%& = & \phi^2\Big(\{\{x_1,x_2\}_P,x_3\}_P+\{\{x_3,x_1\}_P,x_2\}_P+
%\{\{x_2,x_3\}_P,x_1\}_P\Big)\\
%&&+\phi\Big(\{x_1,x_2\}_P\{\phi,x_3\}_P+\{x_3,x_1\}_P\{\phi,x_2\}_P+
%\{x_2,x_3\}_P\{\phi,x_1\}_P\Big)\\
%& = &  \phi\sum_{l=1}^3\frac{\partial\phi}{\partial
%x_l}\Big(\{x_1,x_2\}_P\{x_l,x_3\}_P+\{x_3,x_1\}_P\{x_l,x_2\}_P+
%\{x_2,x_3\}_P\{x_l,x_1\}_P\Big).
%\end{eqnarray*}
%Clearly, for any $l=1,2,3$ the expression under the summation sign
%vanishes.
%\endpf
%

%%%%%%%%%%%%%%%%%%%%%%%%%%%%%%%
%%%%%%%%%%%%%%%%%%%%%%%%%%%%%%%
\subsection{Explicit solutions}
\label{sect: Hirota Euler solutions}
%%%%%%%%%%%%%%%%%%%%%%%%%%%%%%%
%%%%%%%%%%%%%%%%%%%%%%%%%%%%%%%

Explicit solutions were given in \cite{HK}, but it has not been
explained there how to determine the parameters of the elliptic
functions involved in their formulas, using the initial
conditions. We would like to fill in this gap here. We use the
following addition formulas for the Jacobi elliptic functions:
\begin{eqnarray}
{\rm cn}(\xi+\eta)-{\rm cn}(\xi-\eta) & = & -\frac{2\,{\rm
sn}\,\xi\ {\rm dn}\,\xi\ {\rm sn}\,\eta\ {\rm dn}\,\eta}
{1-k^2{\rm sn}^2\xi\ {\rm sn}^2\eta},  \nonumber\\
{\rm sn}(\xi+\eta)-{\rm sn}(\xi-\eta) & = & \frac{2\,{\rm
cn}\,\xi\ {\rm dn}\,\xi\ {\rm sn}\,\eta}
{1-k^2{\rm sn}^2\xi\ {\rm sn}^2\eta},  \nonumber\\
{\rm dn}(\xi+\eta)-{\rm dn}(\xi-\eta) & = &
-\frac{2k^2\,{\rm sn}\,\xi\ {\rm cn}\,\xi\ {\rm sn}\,\eta\
{\rm cn}\,\eta} {1-k^2{\rm sn}^2\xi\ {\rm sn}^2\eta},  \nonumber
\end{eqnarray}
and the related formulas
\begin{eqnarray}
{\rm sn}(\xi+\eta){\rm dn}(\xi-\eta)+{\rm
sn}(\xi-\eta){\rm dn}(\xi+\eta) & = & \frac{2\,{\rm
sn}\,\xi\ {\rm dn}\,\xi\ {\rm cn}\,\eta}
{1-k^2{\rm sn}^2\xi\ {\rm sn}^2\eta},  \nonumber \\
{\rm cn}(\xi+\eta){\rm dn}(\xi-\eta)+{\rm
cn}(\xi-\eta){\rm dn}(\xi+\eta) & = & \frac{2\,{\rm
cn}\,\xi\ {\rm dn}\,\xi\ {\rm cn}\,\eta\ {\rm dn}\,\eta}
{1-k^2{\rm sn}^2\xi\ {\rm sn}^2\eta},\qquad  \nonumber\\
{\rm sn}(\xi+\eta){\rm cn}(\xi-\eta)+{\rm
sn}(\xi-\eta){\rm cn}(\xi+\eta) & = & \frac{2\,{\rm
sn}\,\xi\ {\rm cn}\,\xi\ {\rm dn}\,\eta}{1-k^2{\rm
sn}^2\xi\ {\rm sn}^2\eta}.  \nonumber
\end{eqnarray}
Assume that the coefficients $\delta_i$ are given by formulas
(\ref{delta}) with $\alpha_i$ coming from Eqs. (\ref{alpha for omega})
or (\ref{alpha for M}) with $I_1<I_2<I_3$, so that
\begin{equation}\label{delta signs}
\delta_1<0,\quad \delta_2>0,\quad \delta_3<0.  \nonumber
\end{equation}

Then the above addition formulas suggest to look for the solution
in one of two forms:
\begin{equation}\label{dE sol 1}
 x_1=A_1\,{\rm cn}(\nu n+\varphi_0),\quad
 x_2=A_2\,{\rm sn}(\nu n+\varphi_0),\quad
 x_3=A_3\,{\rm dn}(\nu n+\varphi_0),
\end{equation}
or
\begin{equation}\label{dE sol 2}
 x_1=A_1\,{\rm dn}(\nu n+\varphi_0),\quad
 x_2=A_2\,{\rm sn}(\nu n+\varphi_0),\quad
 x_3=A_3\,{\rm cn}(\nu n+\varphi_0),
\end{equation}
with $\nu$ being a parameter to be determined and $\varphi_0$ an
arbitrary phase. Both possibilities (\ref{dE sol 1}) and (\ref{dE
sol 2}) are realized (in different regions of the phase space).
Consider first the possibility (\ref{dE sol 1}). It is easy to see
that equations of motion (\ref{dE x}) are satisfied by functions
(\ref{dE sol 1}), if and only if the following conditions hold
\cite{HK}:
\begin{eqnarray}
A_1 & = & -\delta_1A_2A_3\,\frac{{\rm cn}(\nu/2)}{{\rm
sn}(\nu/2){\rm dn}(\nu/2)}\,,\label{dE sol 11}\\
A_2 & = & \delta_2A_1A_3\,\frac{{\rm cn}(\nu/2){\rm
dn}(\nu/2)}{{\rm sn}(\nu/2)}\,,\label{dE sol 12}\\
A_3 & = & -\delta_3A_1A_2\,\frac{{\rm dn}(\nu/2)}{k^2{\rm
sn}(\nu/2){\rm cn}(\nu/2)}.\label{dE sol 13}
\end{eqnarray}
The amplitudes $A_i$ should be determined from the values of the
integrals of motion. Substitute the ansatz (\ref{dE sol 1}) into
the integrals (\ref{dE F}), then a direct computation based on the
relations ${\rm cn}^2 \xi =1-{\rm sn}^2 \xi $ and ${\rm dn}^2 \xi
=1-k^2{\rm sn}^2 \xi $ leads to
\[
A_1^2=\frac{1-F_3}{\delta_2\delta_3}\,,\quad
A_2^2=\frac{1-F_3^{-1}}{\delta_1\delta_3}\,,\quad
A_3^2=\frac{1-F_1^{-1}}{\delta_1\delta_2},
\]
and
\[
k^2=\frac{1-F_3^{-1}}{1-F_1}\,.
\]
Thus, this ansatz holds, if and only if $F_1<F_3^{-1}<1$, that is,
if $F_2>1$. With the values just found, relations (\ref{dE sol
11})--(\ref{dE sol 13}) lead to
\[
{\rm sn}^2(\nu/2)=1-F_1.
\]

Turning to the possibility (\ref{dE sol 2}) (omitted in
\cite{HK}), we find that equations of motion (\ref{dE x}) are
satisfied by functions (\ref{dE sol 2}), if and only if the
following conditions hold:
\begin{eqnarray}
A_1 & = & -\delta_1A_2A_3\,\frac{{\rm dn}(\nu/2)}{k^2{\rm
sn}(\nu/2){\rm cn}(\nu/2)}\,,\label{dE sol 21}\\
A_2 & = & \delta_2A_1A_3\,\frac{{\rm cn}(\nu/2){\rm
dn}(\nu/2)}{{\rm sn}(\nu/2)}\,,\label{dE sol 22}\\
A_3 & = & -\delta_3A_1A_2\,\frac{{\rm cn}(\nu/2)}{{\rm
sn}(\nu/2){\rm dn}(\nu/2)}.\label{dE sol 23}
\end{eqnarray}
Substituting the ansatz (\ref{dE sol 2}) into the integrals
(\ref{dE F}), we find:
\[
A_1^2=\frac{1-F_3}{\delta_2\delta_3}\,,\quad
A_2^2=\frac{1-F_1}{\delta_1\delta_3}\,,\quad
A_3^2=\frac{1-F_1^{-1}}{\delta_1\delta_2},
\]
and
\[
k^2=\frac{1-F_1}{1-F_3^{-1}}\,.
\]
Theferore, this ansatz holds, if and only if $F_3^{-1}<F_1<1$,
that is, if $F_2<1$, and then relations (\ref{dE sol
21})--(\ref{dE sol 23}) lead to
\[
{\rm sn}^2(\nu/2)=1-F_3^{-1}.
\]
Thus, in both cases all parameters of the solution are expressed
in terms of the initial data (more precisely, in terms of the
integrals of motion).

%%%%%%%%%%%%%%%%%%%%%%%%%%%%%%%
%%%%%%%%%%%%%%%%%%%%%%%%%%%%%%%
\section{Concluding remarks}
%%%%%%%%%%%%%%%%%%%%%%%%%%%%%%%
%%%%%%%%%%%%%%%%%%%%%%%%%%%%%%%

In this paper, we studied a remarkable birational map of $\bbR^3$,
which serves as an integrable discretization of the Euler top, on
one hand, and plays a role in the spherical geometry, on the
other. Along with a streamlined presentation of results obtained
previously in \cite{J} and in \cite{HK}, namely the conserved
quantities and the solution in terms of elliptic functions, we
found an invariant volume form and a family of compatible
invariant Poisson tensors for this map. Thus, it becomes a
well-established representative of integrable maps, with a
standard definition of integrability in the Liouville-Arnold
sense. One more standard attribute of integrable systems remains
to be found for this map, namely the Lax representation. This
would provide a key to understanding the nature of analogous
discretizations proposed in \cite{KH}, \cite{R}, which seem to
belong to the most mysterious objects in the universe of
integrable maps.

%%%%%%%%%%%%%%%%%%%%%%%%%%%%%%%
%%%%%%%%%%%%%%%%%%%%%%%%%%%%%%%
\section*{Aknowledgments}
%%%%%%%%%%%%%%%%%%%%%%%%%%%%%%%
%%%%%%%%%%%%%%%%%%%%%%%%%%%%%%%

M.P. has been
supported by the European Community through the FP6 Marie
Curie RTN ENIGMA (Contract number MRTN-CT-2004-5652).

%%%%%%%%%%%%%%%%%%%%%%%%%%%%%%%
%%%%%%%%%%%%%%%%%%%%%%%%%%%%%%%

%%%%%%%%%%%%%%%%%%%%%%%%%%%%%%%
%%%%%%%%%%%%%%%%%%%%%%%%%%%%%%%
\end{document}